\documentclass[10pt]{article}
\usepackage{amsmath,amssymb}
\title{CHSH type Bell inequalities involving a party with two or three local binary settings}
\author{Yu-Chun Wu$^{1,2}$, Piotr Badziag$^{1,3}$ and Marek \.Zukowski$^{1,3}$ \\
\small $^1$Institute of Theoretical Physics and Astrophysics, University of Gda\'nsk, PL-80-952 Gda\'nsk\\
\small $^2$Key Laboratory of Quantum Information, \\ \small University of Science and Technology of China, 230026 Hefei, China\\
\small $^3$Stockholm University, Alba Nova University Center, Department of Physics, S106 91, Sweden}
\date{}

\begin{document}

\maketitle

\begin{abstract}
We construct a simple algorithm to generate any CHSH type Bell inequality involving a party with two local binary measurements from two CHSH type inequalities without this party. The algorithm readily generalizes to situations, where the additional observer uses three measurement settings. There, each inequality involving the additional party is constructed from three inequalities with this party excluded. With this generalization at hand, we construct and analyze new symmetric inequalities for four observers and three experimental settings per observer.
\end{abstract}

\section{Introduction}
Not all quantum predictions can be reconciled with the premises of classical local realism. This fact bothered the EPR trio \cite{epr} already in the 1930's. In 1964 John Bell quantified the EPR paradox by constructing an inequality satisfied by all local realistic (classical) correlations but violated by some quantum predictions \cite{bell}. It was the first step toward delineation of the boundaries between Einstein's classical local realism and the genuinely non-classical areas of quantum physics. The task of charting these boundaries has its clear philosophical weight. Moreover, with the current progress in information processing technologies, it is acquiring a utilitarian aspect too. This is because each quantum state exhibiting non-classical correlations can be used as a resource for some distributed information processing \cite{sen03, bruk04}. Thus, there is a strong motivation to construct new tight Bell inequalities for correlations, particularly those with more than two experimental settings per observer. A possible strategy here can be to produce systematic methods of extending simpler inequalities to more parties and/or to more observers. To our knowledge not much has been achieved in this respect so far. 

There is a lifting method \cite{piro}, where the author investigates the Bell polytope made of the Bell experiment outcomes characterized by probability distributions. Properties of the marginal probabilities there can guarantee that the extensions of the Bell inequalities to more observers are valid. However, the resulting extensions are in a sense trivial since they add observers with essentially one experimental setting only. There is also a triangular elimination method  \cite{david} to increase the number of experimental settings per observer without changing the number of observers. Validity of this method is, however, restricted to two observer only.

Here we construct a method to generate a series of 3-settings per observer CHSH-type (for correlations) Bell inequalities and use it to generate and analyze a new class of symmetric 4-observer inequalities. Our method is a direct generalization of a design, which uses 2-settings per observer (c.f., \cite{werwol, zukbru}). Thus we begin the presentation with this simpler setup. There, all the tight inequalities are direct generalizations of the 2-observer Clauser-Horne-Shimony-Holt (CHSH) inequality  \cite{chsh}, which stems from the assumption of local realism and an elementary relation
\begin{equation}
\label{CHSH1}
\frac{a_1+a_2}{2} b_1+\frac{a_1-a_2}{2}b_2 = \pm 1.
\end{equation}
Here we have two observers A and B and each can measure two binary observables $\hat{A}_1,\hat{A}_2$ and $\hat{B}_1,\hat{B}_2$ respectively. The possible values of the measurement outcomes $a_i\ (b_i)$ for observable $\hat{A}_i \ (\hat{B}_i)$ are $\pm 1$. By averaging (\ref{CHSH1}) over a locally realistic distribution of the outcomes, one immediately obtains the CHSH inequality
\begin{equation}
\label{CHSH}| \frac{E_{11} + E_{12} + E_{21} - E_{22}}{2}| \leq 1
\end{equation}
with $E_{ij}$ denoting the average value of $a_i b_j$.

In quantum mechanics binary observables are represented by Pauli matrices. The quantum operator, whose expectation value corresponds to the left-hand side (LHS) of inequality (\ref{CHSH}) is thus
\begin{equation}
\label{Bell2}
\textsl{B}_{2} = \frac{\vec{m}_1+\vec{m}_2}{2} \cdot \vec{\sigma} \otimes \vec{n}_1 \cdot \vec{\sigma} + \frac{\vec{m}_1-\vec{m}_2}{2} \cdot \vec{\sigma} \otimes \vec{n}_2 \cdot \vec{\sigma}.
\end{equation}
The operator norm of the 2-particle Bell operator $\textsl{B}_{2}$ is a function of two angles: between $\vec{m}_1$ and $\vec{m}_2$ and between $\vec{n}_1$ and $\vec{n}_2$. It achieves its maximum of $\sqrt{2}$ when both angles are $\pi/2$. Thus quantum measurements allow for the violation of the CHSH inequality by a factor of $\sqrt{2}$, the Tsirelson bound.

The CHSH inequality was extended to arbitrary number of qubits by Mermin and Ardehali \cite{merin,arde} and finally by Belinskii and Klyshko \cite{belkly}. The n-qubit Bell operators in the generalized Mermin-Ardehali-Belinskii-Klyshko (MABK) inequalities are defined recursively as follows \cite{gisbec}:
\begin{eqnarray}
\nonumber
&&B_1=\vec{a}_1\vec{\sigma}_1, \, B_1'=\vec{a}_1'\vec{\sigma}_1\\
\label{gmabk}
&&B_n=B_{n-1}\otimes \frac12(\vec{a}_n\vec{\sigma}_n+\vec{a}_n'\vec{\sigma}_n)+B_{n-1}'\otimes \frac12(\vec{a}_n\vec{\sigma}_n-\vec{a}_n'\vec{\sigma}_n),
\end{eqnarray}
where $\vec{a}_1(')$ is a real three-dimentional unit vector, $\vec{\sigma}_i=(\sigma^x_i,\sigma_i^y,\sigma_i^z)$ are the Pauli operators for the i-th party and $B_{n-1}'$ can be obtained by exchanging the $\vec{a}_i$ and $\vec{a}_i'$ in the expression for $B_{n-1}$. Inequalities $\langle B_n(\, n\geq2) \rangle \leq 1$ are tight Bell inequalities, which on entangled states can be violated by factor $(\sqrt{2})^{n-1}$. We can show that the Bell operator in every tight CHSH type Bell inequality for correlations between local binary observables and two experimental settings per observer is of form (\ref{gmabk}) and in this case, $B_{n-1}$ and $B_{n-1}'$ may be independent. To facilitate a natural generalization of this statement to the situations, where each observer is allowed three experimental settings, we will prove it by referring to the relation between local realistic correlations and convex polytopes.

\section{Correlation Polytope and its facets}

Local realistic correlations for a given experimental setup form a convex polytope in a statistical hyperspace \cite{pit1}. Each of the facets of the polytope is specified by a linear equation, which in turn corresponds to a tight Bell inequality. The vertices of the polytope represent deterministic predictions (here $\pm 1$), i.e., they specify extremal correlation functions. The extremal functions describe such situations, where the products of the measurement results in every given set of the experimental settings have vanishing standard deviations. Thus, the components of the vertices are $\pm 1$. In particular, when the polytope describes the correlations between the dichotomic observables of many observers, then the vectors representing the vertices are

\begin{equation}
v = \vec{a}\otimes\vec{b} \otimes \ldots =(a_1,a_2,\ldots,a_M)\otimes (b_1,b_2,\ldots,b_N) \otimes \ldots,
\end{equation}
with $a_i$, $b_j$, etc. equal to $\pm 1$. Such vectors are hereafter called
admissible vectors. The full correlation polytope $F_{MN \ldots}$
lives in $M \cdot N \cdot \ldots$ dimensional real space ($M$,  $N$
etc. are the numbers of experimental settings for observers $A,\ B$, etc. respectively). The polytope has inversion symmetry about the origin, so none of its faces may cross the origin. Therefore, every equation
describing a facet can be put in the form
\begin{equation}
\label{facet}
I(\vec{a}, \vec{b}, \ldots) = \sum_{ij \ldots}\alpha_{ij \ldots}a_ib_j \cdot \ldots=1,
\end{equation}
where $\alpha_{ij\ldots}$ are constants.

Given its vertices and the general form of equation (\ref{facet}), it is in general difficult to determine all the facets of the polytope \cite{pit1}, but for arrangements with two experimental settings per observer. With more settings per observer, one can only generate selected families of the facets and their corresponding inequalities \cite{zuk, wbz, ChenDeng08}. 

\section{From facets of $F_{MN}$ to facets $F_{2MN}$}

The main idea behind our design in the next section is visible already when one considers the correlation polytope $F_{2MN}$. Each of its facet is described by an equation like
\begin{equation}
\label{facet2mn}
I(\vec{a},\vec{b},\vec{c}) = a_1 f(\vec{b},\vec{c})+a_2 g(\vec{b},\vec{c}) = 1,
\end{equation}
where $\vec{a}=(a_1,a_2),\vec{b}=(b_1,b_2,\ldots,b_M),\ \vec{c}=(c_1,c_2,\ldots,c_N)$.

The $2MN$ linearly independent vertices, which span the facet, form a set $S=\{\vec{a}_i\otimes \vec{b}_i\otimes\vec{c}_i|i=1,2,\ldots,2MN\}$. Clearly, we can always select the tensor factors in the product so that the first component of $\vec{a}$ is $1$. With this choice it is easy to divide $S$ into two subsets
\begin{equation}
S_1=\{(1,1)\otimes \vec{b}_{1i}\otimes \vec{c}_{1i}\}
\end{equation}
and
\begin{equation}
S_2=\{(1,-1)\otimes \vec{b}_{2i}\otimes \vec{c}_{2i}\}
\end{equation}
where $i=1,2,\ldots$ and $S_1\cap S_2=\varnothing$. Since the whole set $S$ is linearly independent, so are $S_1$ and $S_2$. The same applies to $\{\vec{b}_{1i}\otimes \vec{c}_{1i} \}$ and $\{\vec{b}_{2i}\otimes \vec{c}_{2i} \}$. None of these sets may contain more than $MN$ (linearly independent) elements and together they contain $2MN$ elements (the number of elements in $S$). Thus, there are exactly $MN$ linearly independent vetices which saturate
\begin{equation}
\label{reducedfacet1} I((1,1),\vec{b},\vec{c})=1
\end{equation}
and exactly $MN$ linearly independent vertices saturating
\begin{equation}\label{reducedfacet2}
I((1,-1),\vec{b},\vec{c})=1.
\end{equation}
Since $I(\vec{a},\vec{b},\vec{c})=1$ specifies a facet, then
for any admissible vectors $\vec{b},\vec{c}$ we have
$I((1,1),\vec{b},\vec{c})\leq1$ and
$I((1,-1),\vec{b},\vec{c})\leq1$.

One can thus conclude that every equation (\ref{facet2mn}) describing a facet
of $F_{2MN}$ is generated by two equations describing facets
of $F_{MN}$ in the following way
\begin{equation}
\label{theorem1}
I(\vec{a},\vec{b},\vec{c}) = \frac{a_1 + a_2}{2} I_{+}(\vec{b},\vec{c}) + \frac{a_1 - a_2}{2} I_{-}(\vec{b},\vec{c})=1.
\end{equation}
and
\begin{equation}
\label{extension}
I_{\pm}(\vec{b},\vec{c})= f(\vec{b},\vec{c}) \pm g(\vec{b},\vec{c})=1
\end{equation}
define two facets of $F_{MN}$.

Likewise, every two facets of $F_{MN}$,
$I_{+}(\vec{b},\vec{c})=1$ and
$I_{-}(\vec{b},\vec{c})=1$ generate a facet of
$F_{2MN}$ via equation (\ref{theorem1}).

The claim that every tight CHSH type Bell inequality for n-qubit and
two experimental settings per qubit is of form (\ref{gmabk}) is an immediate
corollary to these statements. 
This is because there is a $1-1$ correspondence between the tight Bell inequalities and the facets of the correlation polytope. A complete set of the facets of $F_{MN}$ generates a complete set of the facets of $F_{2MN}$. Equivalently, a complete set of $M\times N$ CHSH type Bell inequalities yields a complete set of $2\times M\times N$ CHSH type Bell inequalities.

\section{Facets of $F_{3MN}$}

Most of the arguments behind the construction of the 2-setting extensions of tight Bell inequalities readily generalize onto the construction of the 3-setting extensions.

Thus every facet of $F_{3MN}$ is specified by a modification of equation (\ref{facet2mn}),
\begin{equation}
\label{facet3mn}
I(\vec{a},\vec{b},\vec{c}) = a_1 f(\vec{b},\vec{c}) + a_2 g(\vec{b},\vec{c}) + a_3 h(\vec{b},\vec{c}) = 1,
\end{equation}
with $\vec{a}=(a_1,a_2,a_3)$, and the rest of the notation like in equation (\ref{facet2mn}).

Now, the $3MN$ linearly independent vertices, which span the facet, form a set $S=\{\vec{a}_i\otimes \vec{b}_i\otimes\vec{c}_i|i=1,2,\ldots,3MN \}$. This set splits naturally into four non-overlapping subsets.
\begin{eqnarray*}
&S_0=\{(1,1,1)\otimes \vec{b}_{0i}\otimes \vec{c}_{0i}\} = \{ \vec{a}_0 \otimes \vec{u}_{1i}\},\\
&S_2=\{(1,-1,1)\otimes \vec{b}_{2i}\otimes \vec{c}_{2i} \} = \{ \vec{a}_2 \otimes \vec{u}_{2i}\},\\
&S_3=\{(1,1,-1)\otimes \vec{b}_{3i}\otimes \vec{c}_{3i} \} = \{ \vec{a}_3 \otimes \vec{u}_{3i}\},\\
&S_1=\{(1,-1,-1)\otimes \vec{b}_{1i}\otimes \vec{c}_{1i}\} = \{
\vec{a}_1 \otimes \vec{u}_{4i} \}.
\end{eqnarray*}

Unlike previously, vectors $\vec{a}_k$ here are not linearly independent. Thus, one does not need a complete set of vectors $\vec{u}_{ki}$ in every subset $S_k$. Consequently, we cannot claim that every $3MN$ facet is constructed from three or four $MN \ldots$ facets. Nevertheless, one can rewrite equation (\ref{facet3mn}) as
\begin{equation}
\label{facet3mnB}
\frac{a_2 + a_3}{2} I_0(\vec{b},\vec{c}) + \frac{a_1 - a_2}{2} I_2(\vec{b},\vec{c}) + \frac{a_1 - a_3}{2} I_3(\vec{b},\vec{c}) = 1
\end{equation}
with
$I_0 = f+g+h$, $I_2 = f-g+h$ and $I_3 = f+g-h$. We chose this form since $\frac{a_2 + a_3}{2}$ assumes one on $\vec{a}=(1,1,1)$ and vanishes on $\vec{a}=(1,-1,1)$ and on $\vec{a}=(1,1,-1)$. Likewise $\frac{a_1 - a_2}{2}$ is one on $\vec{a}=(1,-1,1)$ and zero on $\vec{a}=(1,1,1)$ and on $\vec{a}=(1,1,-1)$. Finally, $\frac{a_1 - a_3}{2}$ is one on $\vec{a}=(1,1,-1)$ and vanishes on $\vec{a}=(1,1,1)$ and on $\vec{a}=(1,-1,1)$.
One can thus see that when $I_0$,
$I_2$ and $I_3$ describe facets of $F_{MN}$ then there are $3MN$ linearly independent vertices of $F_{3MN}$, satisfying equation (\ref{facet3mnB}). Moreover, for $\vec{a}=(1,1,1)$, $\vec{a}=(1,-1,1)$ and $\vec{a}=(1,1,-1)$, the admissible vertices $\vec{a} \otimes \vec{b} \otimes \vec{c}$ satisfy $\left|I(\vec{a},\vec{b},\vec{c})\right| \leq 1$. Finally, for $\vec{a} = (1,-1,-1)$, one has $I((1,-1,-1),\vec{b},\vec{c})=I_2(\vec{b},\vec{c}) + I_3(\vec{b},\vec{c}) - I_0(\vec{b},\vec{c})$. Thus, an additional requirement that 
\begin{equation}
\label{delta}
I_1(\vec{b},\vec{c})=I_2(\vec{b},\vec{c})+I_3(\vec{b},\vec{c}) -I_0(\vec{b},\vec{c}),
\end{equation}
corresponds to a valid 
inequality (not necessarily tight) guarantees that there are no admissible vertices, for which the left-hand side of (\ref{facet3mnB}) exceeds one. In other words, with these conditions satisfied, equation (\ref{facet3mnB}) describes a facet of $F_{3MN}$ and, consequently, it generates a tight Bell inequality. The equation of the facet can be rewritten as
\begin{equation}
\label{inequality}
a_1\frac{I_2+I_3}2+a_2\frac{I_0-I_2}2+a_3\frac{I_0-I_3}2 = 1
\end{equation}
Here, in contradistinction to the two-setting extensions, one cannot prove that all $3MN$ facets are generated by our extensions. Based on the result of \cite{zuk}, one could hope for such a proof in case of $M=N=3$. A recently published inequality (inequality (5) in \cite{ChenDeng08}), however, successfully nullified such hopes. Thus, for $3$-setting extensions, we can only have a sufficient condition.

\section{A new $3\times 3\times 3\times 3$ Bell inequality}

To illustrate the final result, in this section we construct a four-party tight CHSH type Bell
inequality, which detects some quantum correlations undetectable by the MAKB inequalities. As a departure point we take the $3\times 3\times 3$ inequality found by Wie\'{s}niak et al. \cite{wbz}. There
\begin{eqnarray}
\mathcal{B} &=& \frac{1}{4}\left< \hat{A}_0 (\hat{B}_1 + \hat{B}_2) (\hat{C}_1 - \hat{C}_2) + (\hat{A}_1 - \hat{A}_2) \hat{B}_0 (\hat{C}_1 + \hat{C}_2) +(\hat{A}_1 + \hat{A}_2) (\hat{B}_1 - \hat{B}_2) \hat{C}_0 \right. \nonumber \\
 &+& \left.\frac12 (\hat{A}_1 + \hat{A}_2) (\hat{B}_1 + \hat{B}_2) (\hat{C}_1 + \hat{C}_2) + \frac12 (\hat{A}_1 - \hat{A}_2) (\hat{B}_1 - \hat{B}_2) (\hat{C}_1 - \hat{C}_2)\right> \quad \leq 1,
\end{eqnarray}
where $\hat{A}_i=\vec{A}_i\cdot \vec{\sigma}^A,\,(i=0,1,2)$ denotes the local measurement operators of Alice, $\vec{\sigma}^A$ is the vector consisting of three Pauli matrices  and $\vec{A}_i$ is a real unit vector (likewise for $\hat{B}_i$ and $\hat{C}_i$).  

The most obvious way to satisfy compatibility condition (\ref{delta}) is to choose $I_1=I_2$ or $I_1=I_3$ or $I_3=-I_2$. This is, however, not good since every of these choices leads to a 2-settings extension rather than to a sought for 3-setting extension. One thus needs something more creative. Take, e.g., the following three symmetry transformations on $\mathcal{B}(\vec{a},\vec{b},\vec{c})$:
\begin{enumerate}
\item $x_0\leftrightarrow x_1 \hbox{ and } x_2\leftrightarrow x_2$.
\item $x_0\leftrightarrow x_2 \hbox{  and } x_1\leftrightarrow x_1$.
\item $x_0\rightarrow -x_2 \hbox{ and } x_1\leftrightarrow x_1 \hbox{\quad and } x_2 \rightarrow x_0$.
\end{enumerate}
where $x$ stands for $\vec{A},\ \vec{B}$ or $\vec{C}$, $'\leftrightarrow'$ means swapping the indicated symbols in $\mathcal{B}$; $'\rightarrow'$ means substituting the right part for the left part in the expression for $\mathcal{B}$. These transformations lead to three new tight Bell inequalities generated by $\mathcal{B}_1,\,\mathcal{B}_2,\textrm{ and }\mathcal{B}_3$. Clearly, the new inequalities are equivalent to
the original one. Nevertheless, they represent different facets of $F_{3MN}$. Moreover, it is easy to check that $\mathcal{B}+\mathcal{B}_1=\mathcal{B}_2+\mathcal{B}_3$. Thus, we can put
$I_0=\mathcal{B},\ I_1=\mathcal{B}_1,\ I_2=\mathcal{B}_2,\ I_3=\mathcal{B}_3$ and insert these expressions into inequality (\ref{inequality}) with $\hat{D}_i$ substituting for $a_{i+1}$. The result gives a facet of $F_{3333}$
$$
\hat{D}_0 \frac{\mathcal{B}_2 + \mathcal{B}_3}2 + \hat{D}_1 \frac{\mathcal{B} - \mathcal{B}_2}2 +
 \hat{D}_2 \frac{\mathcal{B} - \mathcal{B}_3}2\leq 1.
$$
With all terms spelled out, it reads
\begin{eqnarray}
\label{fourbell}
\frac18\left\langle \hat{A}_0  [-(\hat{B}_0 - \hat{B}_2) (\hat{C}_0 + \hat{C}_1) (\hat{D}_1 - \hat{D}_2) + (\hat{B}_1 - \hat{B}_0)
            (\hat{C}_0 + \hat{C}_2)(\hat{D}_0 - \hat{D}_1) \right. & \nonumber\\+ (\hat{B}_1 + \hat{B}_2) 
      (\hat{C}_1 - \hat{C}_2) (\hat{D}_1 + \hat{D}_2) + (\hat{B}_0 + \hat{B}_1) (\hat{C}_0 + \hat{C}_1) (\hat{D}_0 -
      \hat{D}_2) &\nonumber\\+ (\hat{B}_0 - \hat{B}_1) (\hat{C}_0 + \hat{C}_2) (\hat{D}_0 - \hat{D}_2)] 
      + \hat{A}_1 [ -(\hat{B}_0 + \hat{B}_2) (\hat{C}_0 - \hat{C}_2) (\hat{D}_1 + \hat{D}_2) &\nonumber\\+ (\hat{B}_1 - \hat{B}_2) (\hat{C}_0 + \hat{C}_1) (\hat{D}_1 -
      \hat{D}_2) + (\hat{B}_0 - \hat{B}_1) 
       (\hat{C}_1 + \hat{C}_2) (\hat{D}_1 - \hat{D}_2) &\nonumber\\+ (\hat{B}_0 + \hat{B}_1) (\hat{C}_0 - \hat{C}_2) (\hat{D}_0 +
      \hat{D}_2) + (\hat{B}_0 + \hat{B}_1) (\hat{C}_1 + \hat{C}_2) (\hat{D}_0 + \hat{D}_2)] \\
      + \hat{A}_2 [ (\hat{B}_0 + \hat{B}_1) (\hat{C}_0 - \hat{C}_1) (\hat{D}_1 - \hat{D}_2) + (\hat{B}_0 - \hat{B}_2) (\hat{C}_0 - \hat{C}_1) (\hat{D}_1 + \hat{D}_2) &\nonumber\\
     \left. + (-\hat{B}_0 + \hat{B}_1) (\hat{C}_0 + \hat{C}_2) (\hat{D}_1 + \hat{D}_2)] \right\rangle  & \leq 1. \nonumber
\end{eqnarray}
To investigate the degree to which this inequality can be violated by measurements on distributed quantum states, one can notice that  
for any two unit vectors  $\vec{Y}\textrm{ and }\vec{Y}'$, $\vec{Y}+\vec{Y}'$ and $\vec{Y}-\vec{Y}'$ are orthogonal and $|\vec{Y}+\vec{Y}'|^2+|\vec{Y}-\vec{Y}'|^2=4$.  Therefore we can always find two orthogonal unit vectors $\vec{X}_i$, with i=1,2, such that $\vec{Y}+\vec{Y}'=2\cos(\alpha)\vec{X}_1\textrm{ and }\vec{Y}-\vec{Y}'=2\sin(\alpha)\vec{X}_2$. Employing this technique, one can choose the local  coordinate axes so that the first two  measurements are parameterized as above and the third is arbitrary. This gives the following parametrization of Alice's measurements (for the other parties, the parametrization will be analogous).
\begin{eqnarray*}\label{substitution}
\hat{A}_0&=&\cos(\chi^A) \hat{X_1^A}+\sin(\chi^A)\hat{X_2^A}\\
\hat{A}_1&=&\cos(\chi^A) \hat{X_1^A}-\sin(\chi^A)\hat{X_2^A}\\
\hat{A}_2&=&\sin(\theta^A)\sin(\phi^A) \hat{X_1^A}+\sin(\theta^A)\cos(\phi^A)\hat{X_2^A}+\cos(\theta^A)\hat{X_3^A}\\
\end{eqnarray*}
where $\{\hat{X_i^A}\}_{i=1}^3$ are a set of local bases chosen by party
Alice. Inserting them into inequality (\ref{fourbell}), and
computing the expectation value for a four-qubit state $\rho$, we
have the inequality:
\begin{equation}\label{cor}
\sum_{i,j,k,l=1}^3 T_{ijkl}\alpha_{ijkl}\leq 4
\end{equation}
where $T_{ijkl}={\bf Tr}(\rho\,
\hat{X_i^A}\otimes\hat{X_j^B}\otimes\hat{X_k^C}\otimes\hat{X_l^D})$ are
the expectation values of the Pauli operators in the local bases for the four-qubit state
$\rho$, and $\alpha_{ijkl}$s are the associated coefficients. The coefficients are combinations of trigonometric functions of $\chi^i,
\theta^i\textrm{ and }\phi^i$ where $i=A,B,C,D$. We can check that
\begin{equation}\label{key}
\sum_{ijkl}\alpha_{ijkl}^2=16.
\end{equation}
This condition together with inequality (\ref{cor}) immediately implies that a four-qubit
state cannot violate inequality (\ref{cor}) as long as
\begin{equation}\label{corcond}
\sum_{ijkl}T_{ijkl}^2 \leq 1.
\end{equation}
holds for any local bases.

We checked the degree of violation of inequality (\ref{fourbell}) on some typical entangled states. The four-qubit GHZ state
$|\Psi\rangle=\frac1{\sqrt{2}}(|0000\rangle+|1111\rangle)$ and  W
state $|W\rangle=\frac12(|0001\rangle+|0010\rangle+|0100\rangle+|1000\rangle)
$ the violations are by factors $2.263$ and $1.448$ respectively. For
the parametric down-conversion (PDC) state \cite{eibl} $
|\Phi\rangle=\sqrt{\frac13}[|0011\rangle+|1100\rangle-\frac12(|0101\rangle-|0110\rangle-|1001\rangle+|1010\rangle)],
$ the violation factor is $1.612$; for the genuine four-qubit entangled
state, which can faithfully teleport an arbitrary two-qubit state
\cite{yeo} $
|\chi\rangle=\frac1{2\sqrt{2}}(|0000\rangle-|0011\rangle-|0101\rangle+|0110\rangle+|1001\rangle+|1010\rangle+|1100\rangle+|1111\rangle)
$ it is $1.579$ and for the four-qubit cluster state in
\cite{brierau} $
|\phi_4\rangle=\frac12(|+\rangle|0\rangle|+\rangle|0\rangle+|+\rangle|0\rangle|-\rangle|1\rangle+|-\rangle|1\rangle|-\rangle|0\rangle+-\rangle|1\rangle|+\rangle|1\rangle)
$ where $|\pm\rangle\equiv (|0\rangle\pm|1\rangle)/\sqrt{2}$, the
violation factor is $1.759$.

Finally, we tested violation of inequality (\ref{fourbell}) on the generalized pure four-qubit GHZ states
\begin{equation}\label{generalghz}
|\Psi\rangle=\cos{(\lambda)}|0000\rangle+\sin{(\lambda)}|1111\rangle,\quad\lambda\in[0,\frac{\pi}4].
\end{equation}
In \cite{scargisin}, it was shown that state (\ref{generalghz}) do not
violates MABK inequality \cite{merin,arde, belkly} if 
$\sin(2\lambda_{cric}) \leq\frac1{\sqrt{8}}$, namely,
$\lambda_{cric} \geq 10.3524^\circ$. With inequality (\ref{fourbell}), the critical value of $\lambda$ is much smaller. According to numerical computation, when $\lambda=1.4324^\circ$  we have a clearly bigger than one
violation factor of $1.001$. Further numerical evidence strongly suggests that inequality (\ref{fourbell}) is violated for all $0 < \lambda < \frac{\pi}{4}$. This observation agrees with the fact that, for state (\ref{generalghz}), $\sum T_{ijkl}^2=5-4\cos{(4\lambda)}$ and this expression is strictly greater than one for all $0 < \lambda < \frac{\pi}{4}$.

In other words, inequality (\ref{fourbell}), although weaker the MAKB inequality, can identify lack of local realism in states with less entanglement than the MAKB inequality. Puzzled by this fact, we looked at the spectra of the corresponding Bell operators. The Bell operator corresponding to the MAKB inequality has two non-degenerate non-zero eigenvalues $\pm 2 \sqrt{2}$. On the other hand, all the eigenvalues of the Bell operator corresponding to inequality (\ref{fourbell}) are non-zero and four of the values ($\pm 2.263$ and $\pm 1.494$) violate the bounds of local realism, $\pm 1$. The remaining eigenvalues are $\pm 0.449$ and $\pm 0.120$ and are triply degenerated. With this fact at hand, it is not all that surprising that our inequality can be violated by the generalized GHZ states only marginally different from a pure product state. Moreover, inequality (\ref{fourbell}) can identify quantum correlations in some relatively strongly mixed states. For instance, an equal mixture of the five states belonging to the three highest eigenvalues there still violates the bound of local realism. By comparison, no equal mixture of three pure orthogonal states can violate a $4$-particle MAKB inequality.

\section{Conclusion}
We have analyzed geometrical relations between the facets of different correlation polytopes and the resulting relations between the equations describing the corresponding facets. This has led us to a method for extending CHSH type Bell inequalities to an increasing number of observers with two or three experimental settings. With two settings allowed to the additional
observer, our method reproduced all the known CHSH type Bell
inequalities for multipartite states. Application of the result to
3-setting extensions, allowed us to construct a new 4-qubit
inequality, which can detect weak quantum correlations,
undetectable by, e.g. the MAKB inequality. 

In principle our scheme can be
applied to construct extensions, where the additional observer is
allowed more than three experimental settings as well. In that case,
however, the scheme may be impractical since the more additional
local settings, the more compatibility  conditions like (\ref{delta}) must be
satisfied. In general, when adding a party with $K$ experimental
settings and constructing a facet of $F_{KMN}$, we need $K$ facets
of $F_{MN}$ and should satisfy $2^{K-1}-K $ compatibility
conditions.

\section{Acknowledgement}

The work is part of EU 6FP programmes QAP and SCALA and has been done at the National Centre for Quantum Information at Gdansk.

\end{document}